\documentclass[12pt,preprint]{aastex}

\shorttitle{Chandra Observations of High Redshift Radio-Loud Quasars}
\shortauthors{LOPEZ ET AL.}

\newcommand{\ltsima}{$\; \buildrel < \over \sim \;$}
\newcommand{\simlt}{\lower.5ex\hbox{\ltsima}}
\newcommand{\gtsima}{$\; \buildrel > \over \sim \;$}
\newcommand{\simgt}{\lower.5ex\hbox{\gtsima}}

\def\lesssim{\mathrel{\hbox{\rlap{\hbox{\lower4pt\hbox{$\sim$}}}\hbox{$<$}}}}
\def\gtrsim{\mathrel{\hbox{\rlap{\hbox{\lower4pt\hbox{$\sim$}}}\hbox{$>$}}}}

\def\arcsec{\hbox{$^{\prime\prime}$}}

\def\aox{$\alpha_{\rm ox}$}

\def\ab1450{$AB_{1450(1+z)}$}

\def\xray{\hbox{X-ray}}

\begin{document}

\title{A Chandra Snapshot Survey of Representative High-Redshift Radio-Loud Quasars from the Parkes-MIT-NRAO Sample}

\author{
L.~A. Lopez,\altaffilmark{1,2,3} W.~N. Brandt,\altaffilmark{1}
C. Vignali,\altaffilmark{4,5} D.~P. Schneider,\altaffilmark{1}
G. Chartas,\altaffilmark{1} and G.~P. Garmire\altaffilmark{1}
}
\altaffiltext{1}{Department of Astronomy and Astrophysics, The
Pennsylvania State University, 525 Davey Laboratory, University
Park, PA 16802, USA; lalopez@astro.psu.edu, niel@astro.psu.edu,
dps@astro.psu.edu, chartas@astro.psu.edu, and garmire@astro.psu.edu.}
\altaffiltext{2}{Institute for Advanced Study, School of Natural Sciences, Einstein Lane, Princeton, NJ 08540}
\altaffiltext{3}{NSF Graduate Research Fellow}
\altaffiltext{4}{Dipartimento di Astronomia, Universita degli Studi di Bologna, Via Ranzani 1, 40127 Bologna, Italy; cristian.vignali@bo.astro.it}
\altaffiltext{5}{INAF --- Osservatorio Astronomico di Bologna, Via Ranzani 1, 40127 Bologna, Italy}

\begin{abstract}
We present the results of {\it Chandra} ACIS-S snapshot
observations of six radio-loud quasars (RLQs) at \hbox{$z \approx$
3.5--4.7}. These observations sample luminous RLQs with
moderate-to-high radio-loudness \hbox{($R \approx$ 200--9600)} and
aim to connect the X-ray properties of radio-quiet quasars \hbox{($R <$ 10)} and
highly radio-loud blazars \hbox{($R >$ 1000)} at high redshift.
This work extends a study by Bassett et al. (2004)
which used similar methods to examine \hbox{$z >$ 4} RLQs with moderate
radio-loudness \hbox{($R \approx$ 40--400)}. All of our targets are clearly detected. A search for extended X-ray emission associated with kpc-scale radio jets revealed only limited evidence for X-ray extension in our sample: three sources showed no evidence of X-ray extension, and the other three had 3--30\% of their total X-ray fluxes extended $>$1\arcsec\ away from their X-ray cores. Additionally, we do not observe any systematic flattening of the optical-to-X-ray spectral index (\hbox{$\alpha_{\rm ox}$}) compared to low-redshift quasars. These results suggest that kpc-scale X-ray jet emission is not dominated by inverse-Compton scattering of CMB-seed photons off jet electrons. We measured X-ray continuum shapes and performed individual and joint spectral fits of our data combined with eight archival RLQs. A single power-law model acceptably fit the data, with best-fit photon indices of 1.72$^{+0.11}_{-0.12}$ for moderate-$R$ sources and 1.47$^{+0.13}_{-0.12}$ for high-$R$ sources. We added an intrinsic absorption component to our model, and neither the moderate-$R$ nor the high-$R$ fits set a lower bound on $N_H$. The upper limits were \hbox{$N_H$ $<$ 3.0 $\times$ 10$^{22}$ cm$^{-2}$} and \hbox{$N_H$ $<$ 2.8 $\times$ 10$^{22}$ cm$^{-2}$}, respectively. Our spectral results suggest that intrinsic absorption does not strongly depend on radio-loudness, and high-$R$ sources have flatter power laws than moderate-$R$ sources. Overall, our high-redshift RLQs have basic X-ray properties consistent with similar RLQs in the local universe.

\end{abstract}


\section{Introduction}

Multiwavelength studies of the high-redshift ($z > 4$) universe have become increasingly widespread over the last decade. With the high sensitivities of X-ray observatories like {\it Chandra} and {\it XMM-Newton}, scientists have analyzed the physical conditions in the immediate vicinities of the first massive black holes to form in the Universe. Prior to 2000, there were only six published X-ray detections of quasars at $z > 4$ (e.g., Kaspi, Brandt, \& Schneider 2000).  In the last five years, the number has risen to over one hundred largely via {\it Chandra} and {\it XMM-Newton} observations.

Wide-area radio surveys, such as the Parkes-MIT-NRAO survey (PMN; Griffith et al. 1995 and references therein), are vital to constrain the properties of high-redshift quasar radio cores, jets, and
lobes. Radio-selected radio-loud quasars (RLQs),
although sampling a minority of the total quasar population, are
less prone to obscuration bias than optically selected targets, as radio
emission is not affected by absorption due to dust.  The
radio-loudness parameter, as defined by Kellermann et al.\ (1989),
is given by \hbox{$R$=\(f_{\rm 5~GHz}/f_{\rm
4400~\mbox{\scriptsize\AA}}\)} (rest frame), where quasars with
$R\ge10$ are RLQs and those with $R<10$ are radio-quiet quasars
(RQQs).

Some of the first X-ray studies of $z>4$ quasars were of a small
group of highly radio-loud blazars (\hbox{$R\approx$~800--27000}) in which the X-ray radiation is
probably dominated by a jet-linked component (e.g., Mathur \& Elvis 1995; Fabian et al. 1997; Moran \& Helfand 1997; Zickgraf et al. 1997). These blazars are not suitable for representative statistical studies, however, because they represent a small minority of even
the RLQ population. Recent X-ray observations with {\it Chandra's} Advanced CCD Imaging Spectrometer (ACIS; e.g., Garmire et al. 2003) have generally focused on $z >$ 4 RQQs (\hbox{$R
\lesssim 2$--10}) which display different \hbox{X-ray} properties than RLQs and thus do not provide much information on RLQ \hbox{X-ray} emission mechanisms. Consequently, understanding of high-redshift RLQ \hbox{X-ray} emission is still limited. Prior to the current paper, there have been ten published detections of $z>4$ RLQs with $R \approx 10$--$1000$, corresponding to typical RLQs in the local universe (excluding the blazar PMN 0525$-$3343 which has $R \approx 800$). 

In comparison to RQQs, RLQs show enhanced \hbox{X-ray} and radio emission that can be attributed to relativistic jet-linked components. For core-dominated, flat-spectrum
RLQs, the majority of the enhanced \hbox{X-ray} emission is presumably from the small-scale (sub-parsec) jet close to the core (e.g., Wilkes \& Elvis 1987; Worrall et al. 1987). In addition,
observations with {\it Chandra} have revealed extended \hbox{X-ray} emission from the kpc-scale jets of RLQs (e.g., Sambruna et al. 2004; Marshall et al. 2005). Several theories
have been developed to explain this unexpectedly strong extended \hbox{X-ray} emission. Low optical luminosities often preclude pure synchrotron emission from a single electron population,
leading some to propose models with more complex electron energy distributions (e.g., Stawarz et al. 2004) or models based on inverse-Compton (IC) scattering. A leading
model for extended X-ray jets invokes bulk relativistic motions (\hbox{$\Gamma \sim$ 1--20}) over kpc scales, thereby allowing electrons in a jet to IC scatter photons from
the Cosmic Microwave Background (CMB) into the X-ray band (e.g., Tavecchio et al. 2000; Celotti, Ghisellini, \& Chiaberge 2001). A natural consequence of the IC/CMB
model is that jets should remain X-ray bright at high redshift: the \hbox{$(1+z)^4$} dependence of CMB energy density compensates for the \hbox{$(1+z)^{-4}$} decrease in surface
brightness (e.g., Schwartz 2002). The IC/CMB model predicts that \hbox{$z > 4$} X-ray jets should often outshine their cores (which dim with luminosity distance squared). As
angular size increases with redshift beyond \hbox{$z \approx 1.6$} in currently favored cosmological models, the sub-arcsecond angular resolution of {\it Chandra} is sufficient to resolve jets with projected length scales $ > $ 10 kpc at all redshifts. Evidence of an X-ray jet has been discovered in the \hbox{$z$ = 4.30} radio-loud (\hbox{$R$ $\approx$ 1200}) quasar \hbox{GB~1508$+$5714} (Siemiginowska et al.\ 2003; Yuan et al.\ 2003). However, the jet-to-core X-ray flux ratio is only $\approx$ 3\%, substantially less than the $\approx$ 100\% predicted by Schwartz (2002). 

In a recent paper (Bassett et al. 2004, B04 hereafter), six new $z>4$ RLQ detections were reported. This study aimed to fill the gap between the many X-ray observations
of $z>4$ RQQs and the extreme blazars. The B04 targets had moderate $R$ parameters of \hbox{$\approx$ 40--400} and were representative of the flat-spectrum RLQ population (Ivezi\'{c} et al. 2002), with several sources near the mode of the $R$-value distribution for RLQs (see Fig.~2 of B04).
B04 found enhanced X-ray emission relative to RQQs of the same redshift and optical luminosity by a factor of $\approx$ 2. Joint spectral analysis revealed the
spectra could be fitted with a power-law model with Galactic absorption and possibly an intrinsic absorption component. The evidence tentatively suggested increased
absorption toward these high-redshift RLQs, generally consistent with previous RLQ studies (e.g., Cappi et al. 1997; Fiore et al. 1998; Reeves \& Turner 2000). No extended
X-ray emission associated with jets from the RLQs was detected, raising possible concern about the IC/CMB explanation of X-ray jets. 

Here we report new {\it Chandra} ACIS-S ``snapshot'' detections of six \hbox{$z\approx$~3.5--4.7} radio-selected RLQs to continue building the observational X-ray
bridge between RQQs and extreme blazars at high redshift. We selected our targets from the large survey of Hook et al. (2002) that searched 7265 deg$^{2}$ for flat-spectrum, radio-loud quasars at high redshift based on a parent sample from the PMN survey. The PMN data cover a significant fraction of the southern sky with \hbox{$\delta$ $<$ 10$^{\circ}$} to a flux-density limit of 20--72 mJy at 5 GHz. Hook et al. found a significant number of new high-redshift RLQs that are bright in the optical and radio wavebands. We chose our targets based on their bright optical and radio fluxes and their high redshifts. Four of our sources were selected to represent the majority population of flat-spectrum RLQs, with moderate-$R$ values ranging 200--1300. Two of our targets are not as representative: PMN~J1230$-$1139 and PMN~J2219$-$2719 are highly radio-loud with \hbox{$R \approx$ 7400--9500}. However, these targets are important as they enlarge the radio-loud ``blazar'' population with X-ray observations (only 5 blazars have been detected in X-rays previously). All the sources have published detections at both 1.4 GHz and 5 GHz and have flat radio spectra with $\alpha_{\rm r} \approx -0.4$~to $+$0.3
($f_{\nu} \propto \nu^{\alpha}$; Hook et al. 2002). We note that our targets have not been mapped at high-resolution in the radio band over multiple epochs to check for, e.g., superluminal motion. All the sources are detected by {\it Chandra} with a minimum of $\approx$~20 counts, as we describe in more detail below. Combined with the results of B04, we have now observed ten moderate-$R$ RLQs at high redshift with {\it Chandra} that are representative of the overall RLQ population. 

This paper brings the total
sample of $z>4$ \xray\ detected RLQs to 21, more than tripling the
number of moderately radio-loud quasars observed before B04 and filling the gap (as
shown in Figure~1) between the many \xray\ observations of $z>4$
RQQs and the extreme blazars.\footnote{
A regularly updated list of $z>4$ quasars with X-ray detections can be
found at
http://www.astro.psu.edu/users/niel/papers/highz-xray-detected.dat}
In combination with the results of B04, this set of quasars is large enough to serve as a representative sample for reliable statistical evaluations of
high-redshift flat-spectrum RLQ \xray\ properties. 

Throughout this paper we adopt $H_{0}$=70 km s$^{-1}$ Mpc$^{-1}$
in a $\Lambda$-cosmology with $\Omega_{\rm M}$=0.3 and
$\Omega_{\Lambda}$=0.7 (e.g., Spergel et al.\ 2003). Additionally, flux densities are implicitly taken to have units of erg cm$^{-2}$ s$^{-1}$ Hz$^{-1}$.

\section{Observations and Data Reduction}
\subsection{Chandra Observations and Basic Data Reduction}

The X-ray observations were obtained by {\it Chandra} during Cycle 5, using the ACIS with the S3 CCD at the aimpoint. The observation log is reported in Table 1. Faint mode was used for the event telemetry format, and {\it ASCA} grade 0, 2, 3, 4, and 6 events were used in the analysis.

Source detection was done using {\sc wavdetect} (Freeman et al. 2002). For each image, we calculated wavelet transforms (using a Mexican-hat kernel) with wavelet scale sizes of 1, 1.4, 2, 2.8, and 4 pixels. Peaks were declared real when their probability of being false was less than the threshold of 10$^{-4}$. This threshold is appropriate when the source position is specified {\it a priori}, as it is here. The {\it Chandra} positions of the detected quasars lie within 0.1--0.3\arcsec~of their optical positions (as reported in Hook et al. 2002; see Table 1), consistent with the expected positional error. Source detection was performed in four energy ranges: the ultrasoft band (\hbox{0.3--0.5~keV}), the soft band (\hbox{0.5--2~keV}), the hard band (\hbox{2--8~keV}) and the
full band (\hbox{0.5--8~keV}). In the redshift range \hbox{$z\approx$~3.5--4.7} for the targets, the full band corresponds to the \hbox{$\approx$~2.5--40}~keV rest-frame band.

The {\sc wavdetect} X-ray photometry is reported in Table 2. The results have been compared to manual-aperture photometry, and good agreement existed between the two methods. Table 2 also shows the hardness ratios [defined as ($H - S$)/($H + S$), where $S$ is the soft-band counts and $H$ is the hard-band counts], the band ratios $(H/S)$, and the effective X-ray power-law photon indices ($\Gamma$) calculated from these band ratios assuming Galactic absorption. The calculaton of $\Gamma$ was performed using the {\it Chandra} X-ray Center Portable, Interactive, Multi-Mission Simulator ({\sc pimms}; Mukai 2002) software. A time-dependent correction was applied for the quantum-efficiency degradation of {\it Chandra} ACIS at low energies using {\sc ciao} Version 3.2.2. All of the quasars have $\Gamma$ values consistent, within the significant errors, with previous studies of RLQs at $z\approx$~0--2 ($\Gamma \approx$~1.4--1.9; 
e.g., Reeves \& Turner 2000).  Basic full-band spectral
analyses were also performed for all of the targets using the Cash statistic (Cash
1979), a technique appropriate for low-count sources (e.g., Nousek \& Shue 1989), with {\sc xspec} (Version~11.2.0; Arnaud et al.\ 1996). We found general agreement between the band-ratio and Cash-statistic
methods. 

Figure~2 shows the soft-band images (chosen to give the best signal-to-noise
ratio) of the six RLQs with raw 0.492$\arcsec$ pixels.

\subsection{Companion X-ray Sources}

We searched for possible companion sources and detached jets over a region of \hbox{$100\arcsec \times 100\arcsec$} centered on each quasar. At \hbox{$z \approx 3.5$--$4.7$}, 100\arcsec\ corresponds to a linear scale of \hbox{$\approx$ $730$--$650$} kpc. In the six fields, we find a total of two possible companion sources in the soft band, one each in the fields of PMN J0235$-$1805 and PMN J1230$-$1139. Neither of these is sufficiently \hbox{X-ray} bright or close to our targets to contaminate their \hbox{X-ray} emission. Both of these sources have $B$-band counterparts in the second Palomar Optical Sky Survey (POSS II) plates, which likely precludes them from being companions to the quasars at high redshift. 

As a further rough observational check, we compared the \hbox{0.5--2~keV} cumulative
number counts of the sources detected in the quasar fields (the
targeted objects were, of course, excluded) with those presented
by Moretti et al.\ (2003) using a compilation of six different
surveys.  The surface density of \xray\ sources at our observed
flux limit (\hbox{$\approx 2\times 10^{-15}$~erg cm$^{-2}$ s$^{-1}$}) is
\hbox{$N(>S) = 432^{+569}_{-279}$~deg$^{-2}$}, consistent within the
errors with the results (490~deg$^{-2}$ at this flux limit) of
 Moretti et al.\ (2003).  We find no evidence for an excess of X-ray sources associated with our quasars.

\subsection{X-ray Extension of the Quasars}

We used the same method as B04 to constrain the presence of either X-ray jets close to the sources or gravitational lensing (e.g., Wyithe \& Loeb 2002). We performed an
analysis of \xray\ extension for all the quasars in the present
sample.  For each quasar, we created a point spread function (PSF)
at \hbox{$\approx$~1.5} keV at the source position (using the {\sc ciao} tool {\sc mkpsf}) normalized by the observed number of source counts.
The radial profile of this PSF was compared with that of the
source.  To identify a putative jet, we required a minimum of three nearby counts offset
by $>1\arcsec$ from the core, corresponding to a distance
of $>8$~kpc at $z=3.5$--4.7. 

Using this method, three of the targets (PMN~J0214$-$0518, PMN~J0324$-$2918, and PMN~J1230$-$1139) showed no evidence of X-ray extension. The other three showed slight signs of extended X-ray emission based on our PSF analysis or on visual inspection of the raw images. PSF analysis of the PMN~J0235$-$1805 X-ray snapshot revealed an excess at $\approx$2\arcsec\ from the core, and seven counts are extended toward the Western direction in the raw X-ray image ($\approx$30\% of the total X-ray flux from the source; see Fig. 2). PSF analysis of the PMN~J1451$-$1512 and PMN~J2219$-$2719 snapshots displayed a slight excess of counts between 1 and 2 arcseconds from the core. The raw X-ray image of PMN~J2219$-$2719 showed six counts extended to the South ($\approx$3\% of the total X-ray flux from the source) which are coincident with extended radio emission observed from the source (C. C. Cheung et al., in preparation). From these results and those of B04, we constrain any spatially resolvable jets to be $\approx$ 3--30 times fainter than their X-ray cores. It was not possible to overlay high-resolution radio contours on Figure 2. The highest resolution radio data available to us are from the NVSS (Condon et al. 1998), and the corresponding $\approx$ 45\arcsec FWHM (Condon et al. 1998) resolution is insufficient to reveal any relevant extended radio structure. 

\subsection{X-ray Variability}

To search for X-ray variability of our sources, we analyzed the photon arrival times in the full band
using the Kolmogorov-Smirnov (KS) test. No significant variability
was detected. We note that this analysis of X-ray variability for the
quasars is limited by both the short X-ray exposures
in the rest frame (\hbox{$\approx$~15~min}; see Table~1) and the relatively small number of counts for most of our quasars.

\section{X-ray Analysis}

The principal X-ray, optical, and radio properties of the six RLQs from our sample are given in Table 3. The columns are as follows:

\noindent
{\sl Column (1)}. --- The name of the source. \\
{\sl Column (2)}. --- The redshift of the source. \\
{\sl Column (3)}. --- The Galactic column density (from Dickey \& Lockman 1990) in units of \hbox{10$^{20}$ cm$^{-2}$}. \\
{\sl Column (4)}. --- The monochromatic rest-frame \ab1450\
magnitude (defined in $\S$3b of Schneider et al.\ 1989), corrected
for Galactic extinction. To calculate \ab1450\, we used $R$-band magnitudes from the APM Catalog
(McMahon et al.\ 2002) and the empirical relationship
\hbox{$AB_{1450(1+z)}=R-(0.648)z+3.10$}. This relationship provides reliable $AB_{1450(1+z)}$ estimates
(within \hbox{$\approx$~0.1--0.2} mags)
in the redshift range under consideration.\\
{\sl Columns (5) and (6)}. --- The 2500~\AA\ rest-frame flux
density and luminosity. These were computed from the \ab1450\
magnitude assuming an optical power-law slope of $\alpha=-0.5$
\hbox{($f_{\nu}$ $\propto$ $\nu^{\alpha}$}; Vanden Berk et al.\
2001). The 2500~\AA\ rest-frame flux densities and luminosities
are increased by \hbox{$\approx$~15\%} for an optical power-law slope of
\hbox{$\alpha=-0.79$} (e.g., Fan et al.\ 2001)
as in Vignali et al.\ (2001, 2003a,b). \\
{\sl Column (7)}. --- The absolute $B$-band magnitude computed
assuming \hbox{$\alpha=-0.5$}. If $\alpha=-0.79$ is adopted for
the extrapolation,
the absolute $B$-band magnitudes are brighter by \hbox{$\approx$~0.35} mag. \\
{\sl Columns (8) and (9)}. --- The observed count rate in the
0.5--2~keV band and the
corresponding flux ($f_{\rm X}$), corrected for Galactic
absorption. This flux has been calculated using {\sc pimms} and a
power-law model with \hbox{$\Gamma=1.7$}. Changes of the photon index in the range
\hbox{$\Gamma=1.4$--1.9} lead to only a few percent change in the
measured \xray\ flux. The \xray\ fluxes reported and used in this paper have been
corrected for the ACIS quantum-efficiency degradation at low energy using
the method described in $\S3.1$. \\
{\sl Columns (10) and (11)}. --- The rest-frame 2~keV flux density and luminosity, computed assuming \hbox{$\Gamma=1.7$}. \\
{\sl Column (12)}. ---  The \hbox{2--10~keV} rest-frame luminosity, corrected for Galactic absorption. \\
{\sl Column (13)}. ---  The optical-to-X-ray power-law slope, \aox, defined as
\begin{equation}
\alpha_{\rm ox}=\frac{\log(f_{\rm 2~keV}/f_{2500~\mbox{\rm \scriptsize\AA}})}{\log(\nu_{\rm 2~keV}/\nu_{2500~\mbox{\rm \scriptsize\AA}})}
\end{equation}
where \hbox{$f_{\rm 2~keV}$} and $f_{2500~\mbox{\scriptsize \rm \AA}}$
are the rest-frame flux densities at 2~keV and 2500~\AA,
respectively. The \hbox{$\approx 1\sigma$} errors on \aox\ have been
computed following the ``numerical method'' described in
$\S$~1.7.3 of Lyons (1991). Both the statistical uncertainties on
the \xray\ count rates and the effects of possible changes in the
\xray\ (\hbox{$\Gamma\approx$~1.4--1.9}) and optical
(\hbox{$\alpha\approx$~$-$0.5} to $-$0.9; Schneider et al. 2001)
continuum shapes have been taken into account. \\
{\sl Column (14)}. --- The radio power-law slope \hbox{($f_{\nu}$
$\propto$ $\nu^{\alpha}$)} between 1.4~GHz and 5~GHz (observed
frame) from Hook et al. (2002). We note that the objects in our sample, particularly those with high $R$ values, may be variable, affecting our calculated values of $\alpha_r$. \\
{\sl Column (15)}. --- The radio-loudness parameter defined by
Kellermann et al.\ (1989) as \hbox{$R=f_{\rm 5~GHz}/f_{\rm 4400~\mbox{\scriptsize\AA}}$} (rest frame). The rest-frame 5~GHz
flux density was computed from the NVSS observed 1.4~GHz flux density and the radio power-law slope given in column 14. The rest-frame 4400~\AA\ flux density was computed from the \ab1450\ magnitude assuming an optical power-law slope of \hbox{$\alpha=-0.5$}. Variability of our targets may affect our measurements of $R$.  \\

\subsection{Joint Spectral Fitting}

We constrained the average X-ray spectral properties of high-redshift flat-spectrum RLQs by performing joint spectral analyses using fourteen RLQs observed by {\it Chandra}, our sample combined with eight RLQs ($z$ $\approx$ 4.1--5.1) from Vignali et al. (2001) [\hbox{SDSS 0210$-$0018}], Vignali et al. (2003a) [\hbox{PSS 0121$+$0347}], Vignali et al. (2003b) [\hbox{SDSS 0913$+$5919}], and B04 [\hbox{FIRST 0725$+$3705}, \hbox{SDSS 0839$+$5112}, \hbox{CLASS J1325$+$1123}, \hbox{FIRST 1423$+$2241}, and \hbox{CLASS J0918$+$0636}]. Source counts were extracted from 2\arcsec\ radius circular apertures centered on the X-ray position of each quasar. The background was taken from annuli centered on the targets such that they avoided the presence of nearby faint X-ray sources. The 14 RLQs do not appear to be biased by the presence of objects with atypically high signal-to-noise ratios, and the removal of any one target does not produce significantly different results. 

We corrected our spectral responses for the quantum-efficiency degradation of ACIS at low energies using the {\it Chandra} X-ray Center's time-dependent calibration given by the {\it Chandra} Calibration Database ({\sc caldb}) Version 3.1.0. Although degradation is not severe above \hbox{$\approx$ 0.7} keV, uncorrected responses could produce incorrect estimates of X-ray absorption. Joint spectral fitting of the unbinned quasar spectra was performed using the Cash statistic. By utilizing the Cash statistic with unbinned data, all the available spectral information was retained. Data below \hbox{0.3 keV} and above \hbox{8 keV} were ignored. 

We divided the 14 RLQs into two groups based on their $R$ values, one with \hbox{$R$ $\approx$ 80--380} (eleven sources with a median value $R$ $\approx$ 220 and $\approx$ 500 total counts) and another with \hbox{$R \approx$ 1280--9550} (three sources with a median value $R$ $\approx$ 7400 and $\approx$ 400 total counts). We performed joint spectral fits of each group to assess the X-ray spectral properties of moderate-$R$ and high-$R$ RLQs separately, and all errors were calculated using \hbox{$\Delta C$ = 2.71}. We first fit a power-law model with each source assigned its own Galactic absorption and redshift. The best-fit photon index for the moderate-$R$ group was 1.72$^{+0.11}_{-0.12}$; the best-fit photon index for the high-$R$ group was 1.47$^{+0.13}_{-0.12}$. The fitted photon indices are consistent with those derived using the band ratios of all 14 targets, from which we obtain a weighted mean of $\Gamma$ = 1.71$\pm$0.11 for the moderate-$R$ group and $\Gamma$ = 1.40$\pm$0.13 for the high-$R$ group. The results are also consistent with previous studies of $z$ $\approx$ 0--4 RLQs (e.g., Brinkmann et al. 1997, Reeves \& Turner 2000; B04), which found core-dominated, flat-spectrum RLQs typically have $\Gamma$ $\approx$ 1.4--1.9. 

To search for evidence of intrinsic X-ray absorption in the two groups, we fit a power-law model with Galactic absorption as well as a redshifted neutral-absorption component (the {\sc xpsec} model zphabs). The confidence contours in the $N_H$-$\Gamma$ plane are shown in \hbox{Figure 3}. For the moderate-$R$ subsample, we obtained a best-fit photon index of $\Gamma$ = 1.80$\pm$0.16, consistent with $\Gamma$ values in Table 2. The 68\% confidence region does not set a lower limit on the intrinsic column density, indicating the possibility of no intrinsic absorption. The upper limit is \hbox{$N_H$ $<$ 3.0 $\times$ 10$^{22}$ cm$^{-2}$} (\hbox{90\%} confidence). 

The analysis of the high-$R$ sample produced similar results. We obtained a best-fit photon index of $\Gamma$ = 1.57$\pm$0.18 and an upper limit on the intrinsic column density of \hbox{$N_H$ $<$ 2.8 $\times$ 10$^{22}$ cm$^{-2}$} (\hbox{90 \%} confidence). Comparison of the low-$R$ and high-$R$ results does not reveal any strong dependence of $N_H$ on radio loudness. Our data are highly suggestive that high-$R$ sources have flatter photon indices than moderate-$R$ sources, consistent with low-redshift observations of RLQs. Tighter spectral constraints in future observations can further test these conclusions.

Earlier studies of RLQs (e.g., Cappi et al. 1997; Fiore et al. 1998; Reeves \& Turner 2000; Page et al. 2005) suggest a significant $N_H$-$z$ correlation, whereby the fraction of RLQs with X-ray absorption seems to increase with redshift: $z \approx$ 2$-$4 RLQs have typical column densities of a few \hbox{$\times 10^{22}$ cm$^{-2}$}. These columns are within the confidence limits of our joint spectral fits for both the moderate-$R$ and the high-$R$ groups. 

To assess the effects of the {\it Chandra} ACIS low-energy contamination on our results, we performed additional joint spectral fits of the two groups using only data from 0.5--8 keV. The results of this analysis yield best-fit values for $N_H$ slightly less than those above, but the results are consistent within the limits. We set an upper limit on the average column density of \hbox{$N_H$ $<$ 3.2 $\times$ 10$^{22}$ cm$^{-2}$} (90 \% confidence) for the moderate-$R$ sources and \hbox{$N_H$ $<$ 3.9 $\times$ 10$^{22}$ cm$^{-2}$} (90 \% confidence) for the high-$R$ sources.  Similar to the results above, the 68 \% confidence region does not set a lower limit on the intrinsic $N_H$. 

\subsection{Nature of the Jet-Linked Component}

Our high-redshift RLQ targets are $\approx$ 3--20 times more \hbox{X-ray} bright than RQQs with comparable optical luminosities and redshifts (see Figure 4). The enhanced X-ray emission of RLQs in comparison to RQQs is usually attributed to a sub-pc relativistic jet-linked component. On the kpc-scale, bright X-ray jets have been observed in $z \approx$ 0.01--2.11 RLQs with {\it Chandra} that some attribute to IC scattering of CMB-seed photons off jet electrons. However, we do not observe bright kpc-scale X-ray jets in our sample and that of B04, counter to the predictions of the IC/CMB model (see $\S1$): nine of the twelve RLQs do not show any evidence of X-ray extension. The other three showed only slight signs of extended X-ray emission, with 3--30\% of their total X-ray flux originating from $>$1\arcsec\ of their X-ray cores (see $\S$2.3 and Fig. 2). The projected length scales for the detected X-ray extension range from 5--10 kpc and do not obviously depend on the radio-loudness of the source. With the exception of PMN~J1230$-$1139, we note that the X-ray, radio, and optical cores of each quasar are coincident (within 0.21\arcsec), indicating the observed X-ray emission is not primarily from large-scale jets. In the case of PMN~J1230$-$1139, the larger offset of the \hbox{X-ray} and optical cores (0.35\arcsec ) may be attributed to inaccuracy in the optical astrometry. To confirm this assessment, we considered the 4-cm and 13-cm images of PMN~J1230$-$1139 taken for the Very Long Baseline Array Calibrator Survey (VCS1; Beasley et al. 2002). The flux seen in these images appears almost certain to be from the radio core, given its morphology and intensity (A.~J. Beasley and D.~Dallacasa 2005, private communication). The VLBA core position is consistent to within 0.26\arcsec\ with the X-ray core (see Table 1). Thus, we conclude the X-ray emission originates from the quasar core and is not from extended X-ray jets. 

In Figure 5 we have plotted the broad-band spectral index $\alpha_{\rm ox}$ versus radio loudness for our high-redshift RLQs (triangles) and samples of radio-selected flat-spectrum RLQs (Brinkmann et al. 1997) which span $z < 2$ (filled circles) and $z$ = 2--4 (open circles). Schwartz (2002) predicts that jets at $z>4$ should be as bright or brighter in X-rays than their quasar cores. Such an enhancement would lead to a systematic flattening of $\alpha_{\rm ox}$ by $\approx$0.1--0.2 relative to comparable objects at low redshift. We do not observe this systematic flattening, further indicating that the IC/CMB process does not dominate the total X-ray production within all flat-spectrum RLQs. 

In combination with the results of B04, we now have a respectably sized sample for reliable statistical evaluations of high-redshift flat-spectrum RLQ \xray\ properties. The larger sample reduces sensitivity to jet-orientation effects. It is possible that our two high-$R$ sources (PMN~J1230$-$1139 and PMN~J2219$-$2719) are preferentially observed extremely close to the line-of-sight of their jet axes (and thus show little-to-no spatial extension). However, it is statistically unlikely that all ten of our moderate-$R$ targets from this paper and B04 would have similar jet orientation. Note that the $R$ values of these ten sources ($\approx$ 40--1200) are much lower than the $R$ values observed for highly-aligned blazars (that have a median $R$ value of $\approx$ 8,000) \footnote{As a further check on this analysis, we compared our source positions with the positional error contours of unidentified EGRET sources (Hartman et al. 1999). We found that none of these contours is consistent with the positions of targets from this paper or B04, supporting the conclusion that none of our sources is a highly-aligned blazar.}. The complete sample from this paper and B04 (moderate-$R$ and high-$R$ sources combined) now spans $R$ $\approx$ 100--9600, encompassing the radio-loudness range of both \hbox{3C 273} ($R$ $\approx$ 1500) and PKS 0637$-$752 ($R$ $\approx$ 2200) which have clearly extended X-ray jets and are not highly-aligned blazars. Schwartz (2002) uses jets from the quasars \hbox{3C 273} (\hbox{$z$ = 0.158}) and \hbox{PKS 0637$-$752} (\hbox{$z$ = 0.654}) as examples of IC/CMB jets (at projected distances of roughly 25 and 75 kpc, respectively) that should outshine their parent quasars above \hbox{$z$ $\approx$ 4}. Both of these quasars have flat (core) radio spectra similar to those of our objects, and the predicted {\it Chandra} count rates of the cores as they would appear at \hbox{$z$ $\approx$ 4} (see Fig. 2 of Schwartz 2002) of a few times 10$^{-3}$ counts s$^{-1}$ are consistent with our observations. The excellent agreement of the optical and X-ray core positions in Figure 2, the absence of X-ray extension observed on scales above $\approx$ 8 kpc, and the lack of systematic $\alpha_{\rm ox}$ flattening of our targets compared to low redshift RLQs indicates the majority of high-redshift RLQ X-ray emission originates from their X-ray cores and not from kpc-scale X-ray jets. Our results contrast with the predictions of the IC/CMB model, showing that any spatially resolvable X-ray jets must be $\approx$ 3--30 times fainter than their X-ray cores. 

On sub-parsec scales, X-rays are probably produced primarily where the quasar photon field dominates the CMB photon field. On kpc scales, X-ray jets in RLQs may be produced via synchrotron radiation by multiple electron populations (e.g., Atoyan \& Dermer, 2004; Stawartz et al. 2004). Alternatively, RLQ jets could perhaps be stifled at high redshift by their environments and only rarely achieve large angular sizes. Recent radio work on PMN~J2219$-$2719 (C.~C. Cheung et al., in preparation) shows that jets cannot be stifled totally as the source has a prominent radio jet coincident with extended X-ray emission. An alternate possibility is that the jets are decelerated via injection of external material (e.g., Hardcastle et al. 2006). If the jet is composed of e$^{+}$-e$^{-}$ pairs, it is possible the high-$z$ CMB-photon energy density may be sufficient to decelerate the jets on the observed kpc scales, although the properties of the jet particles need to be constrained further to test this theory. Future resolved X-ray and radio imaging of $z > 4$ flat-spectrum RLQs is necessary to test these hypotheses and determine the mechanisms responsible for RLQ X-ray extension. 

\section{Summary and Future Work}

Our primary results are the following:

1. We have presented new {\it Chandra} observations of six RLQs at $z \approx$ 3.5--4.7, significantly increasing the number of X-ray detected, high-redshift RLQs. Our targets have moderate-to-high radio-loudness parameters, with $R$ $\approx$ 200--9600. 

2. We do not detect significant extended X-ray emission associated with kpc-scale X-ray jets. Nine RLQs from our sample and that of B04 showed no X-ray extension, and three RLQs displayed slight extension with only 3--30\% of their total X-ray fluxes extending $>$1\arcsec\ away from their X-ray cores. This result is on the order of the modest X-ray jet discovered in the $z$ = 4.3 blazar GB 1508$+$5714  (where $\approx$ 3\% of the total flux was extended from the core). We conclude any spatially resolvable jets must be at 3--30 times fainter than their X-ray cores. Additionally, we do not observe any systematic flattening of optical-to-X-ray spectral index compared to low-redshift quasars. In combination with the results of B04, we have a large enough sample to reduce sensitivity to jet-orientation effects. Our results contrast with the predictions of the IC/CMB model, and we conclude that kpc-scale X-ray production is not dominated by a straightforward IC/CMB process in flat-spectrum RLQs. 

3. Individual and joint spectral fitting revealed 14 RLQs can be fit acceptably using a power-law model with a Galactic absorption component. The average best-fit photon indices of \hbox{$\Gamma$ = 1.72$^{+0.11}_{-0.12}$} for the moderate-$R$ objects and \hbox{$\Gamma$=1.47$^{+0.13}_{-0.12}$} for the high-$R$ objects are consistent with similar objects at low redshift. To search for evidence of intrinsic X-ray absorption, we added a redshifted neutral-absorption component to the above model. Neither the moderate-$R$ nor the high-$R$ fits set a lower bound on $N_H$: the upper limits were \hbox{$N_H$ $<$ 3.0 $\times$ 10$^{22}$ cm$^{-2}$} and \hbox{$N_H$ $<$ 2.8 $\times$ 10$^{22}$ cm$^{-2}$}, respectively. Our results are consistent (within the 90 \% confidence limits) with previous quasar and blazar studies. Our results indicate intrinsic absorption $N_H$ does not have a strong dependence on radio-loudness. 

Longer X-ray exposures with {\it Chandra} and {\it XMM-Newton} of our sample are necessary to investigate further high-redshift RLQ X-ray extension, X-ray absorption, and X-ray variability. The greater photon statistics of deeper X-ray observations will enable direct detection of X-ray jets as well as provide tighter constraints on intrinsic X-ray absorption. Additionally, longer X-ray exposures will yield better timescales to explore RLQ \hbox{X-ray} variability and its dependence on redshift. High-resolution radio imaging of high-redshift RLQs can reveal the direction and magnitude of extended radio emission, and these results can be compared to extended X-ray features to test for possible coincidence and to verify putative jets. 

\acknowledgements

We thank L.~C. Bassett, A.~J. Beasley, C.~C. Cheung, D. Dallacasa, and A.~T. Steffen for helpful discussions. This work was supported by a NASA grant NAS8-01128 (GPG, Principal Investigator), National Science Foundation Graduate Research Fellowship (LAL), NASA LTSA grant NAG5-13035 (LAL, WNB, DPS), NSF Grant AST03-07582 (DPS), and MIUR COFIN grant 03-02-23 (CV).

\clearpage
\begin{figure*}[p]
\figurenum{1}
\centerline{\includegraphics[angle=0,width=15cm]{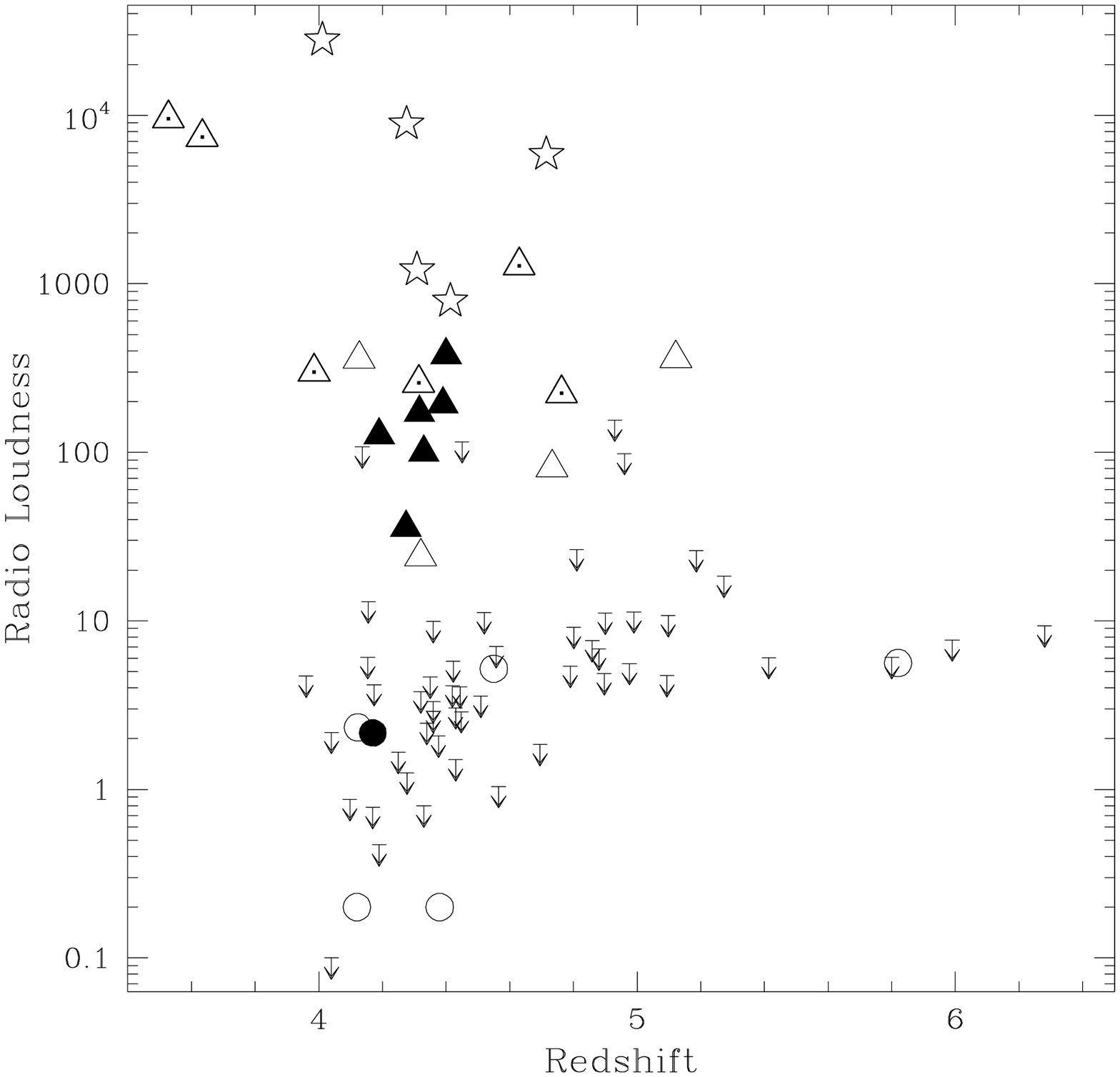}}
\figcaption[f1.eps]{Radio loudness versus redshift for X-ray detected quasars at $z\geq 4$. In addition, quasars from this paper between $3.5 < z < 4.0$ are included. The large symbols represent RLQs ({\it triangles}), blazars ({\it stars}), and RQQs ({\it circles}).  The filled symbols represent the quasars in the B04 sample, and the triangles with central dots represent the targets from the current sample. We tentatively classify the high-$R$ targets in our sample as RLQs, although they may be considered blazars. Arrows show upper limits
for radio loudness at the $\approx 3\sigma$~level based on the
1.4~GHz radio flux density. Note that, excluding
upper limits, the present and B04 samples more than triple the number of
objects with moderate radio-loudness (\hbox{$R \approx$~10--1000}),
filling the X-ray observational gap between the RQQs and the
blazars.}
\end{figure*}
\begin{figure*}[p]
\figurenum{2}
\centerline{\includegraphics[angle=0,width=17cm]{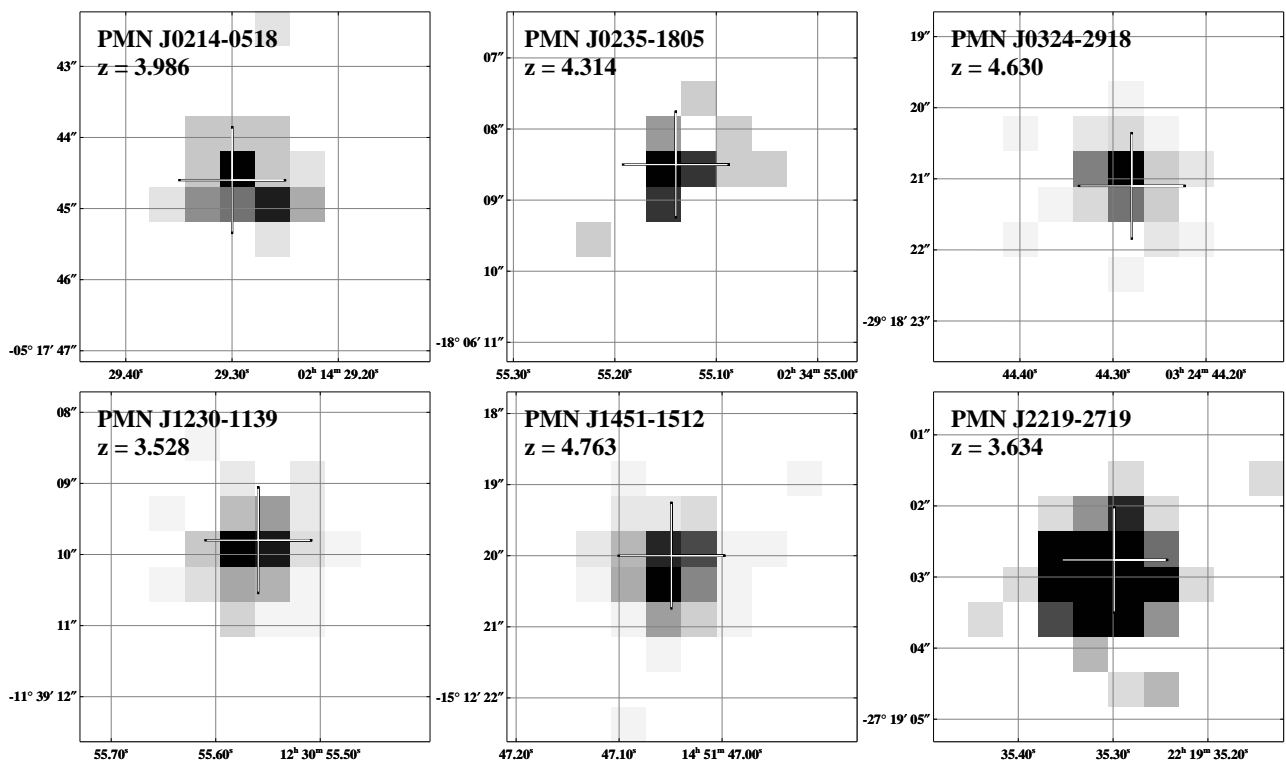}}
\figcaption[f2.eps]{{\it Chandra} 0.5--2 keV ($\approx$ 2.5--10 keV rest-frame) images of the six RLQs presented in this paper. Images are 5\arcsec\ $\times$ 5\arcsec\, showing raw 0.492\arcsec\ pixels. In each panel the horizontal axis shows the right ascension, and the vertical axis shows the declination (both in J2000.0 coordinates). North is up, and East is to the left. Crosses mark the optical positions of the quasars (the cross marks the radio core position in the case of PMN~J1230$-$1139).}
\end{figure*}
\begin{figure*}[p]
\figurenum{3}
\centerline{\includegraphics[angle=0,width=10cm]{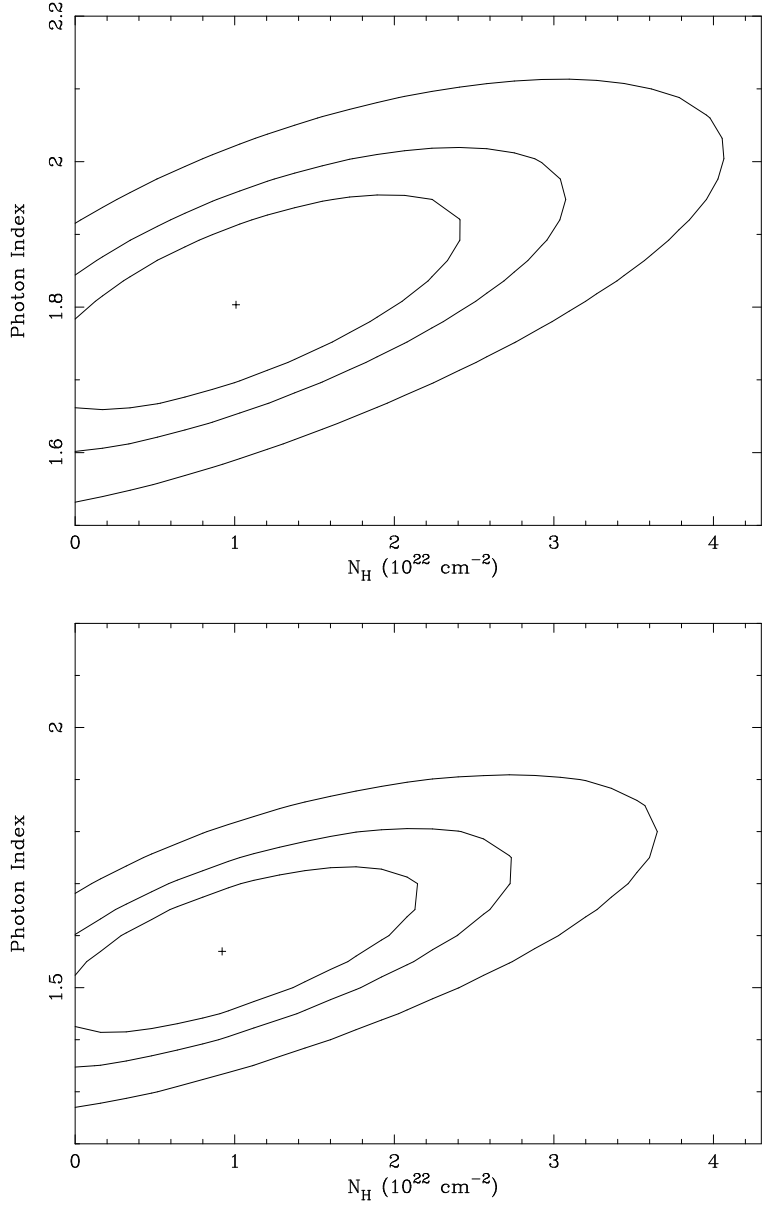}}
\figcaption[f3.eps]{{\bf Top}: $N_H$-$\Gamma$ confidence contours for the eleven moderate-$R$ quasars in our sample. Inner, middle, and outer contours represent 68\%, 90\%, and 99\% confidence regions, respectively. The 68 \% confidence region does not set a lower limit on the intrinsic $N_H$ indicating the possibility of no intrinsic absorption. {\bf Bottom}: $N_H$-$\Gamma$ confidence contour for the three high-$R$ quasars in our sample. Similar to the moderate-$R$ group, the fit did not produce a lower bound on $N_H$.}
\end{figure*}
\begin{figure*}[p]
\figurenum{4}
\centerline{\includegraphics[angle=0,width=15cm]{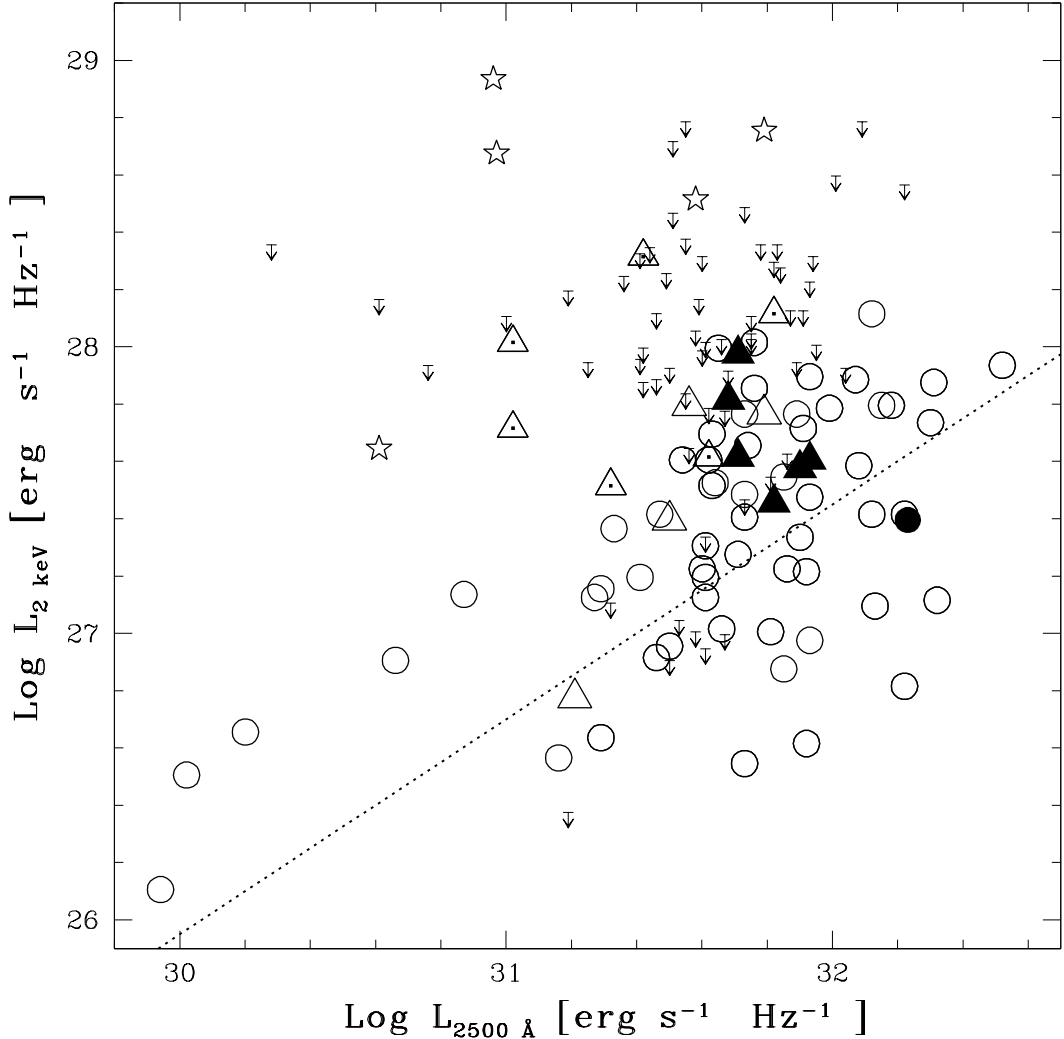}}
\figcaption[f4.eps]{X-ray (2 keV) vs. UV (2500 \AA) monochromatic luminosity for $z$ $>$ 4 RQQs ({\it circles}), RLQs ({\it triangles}), blazars ({\it stars}), and X-ray upper limits ({\it downward-pointing arrows}, at the \hbox{$\approx$3 $\sigma$} confidence level). The filled symbols represent the quasars in the B04 sample, and the triangles with central dots represent the targets from the current sample (\hbox{3.5 $<$ $z$ $<$ 4} quasars from the current sample are included). The dotted line indicates the best-fit relationship for 137 SDSS RQQs in the 0.16--6.28 redshift range (Vignali et al. 2003a). The current sample is $\approx$ 3--20 times X-ray brighter than RQQs of comparable UV luminosities.}
\end{figure*}
\begin{figure*}[p]
\figurenum{5}
\centerline{\includegraphics[width=15cm]{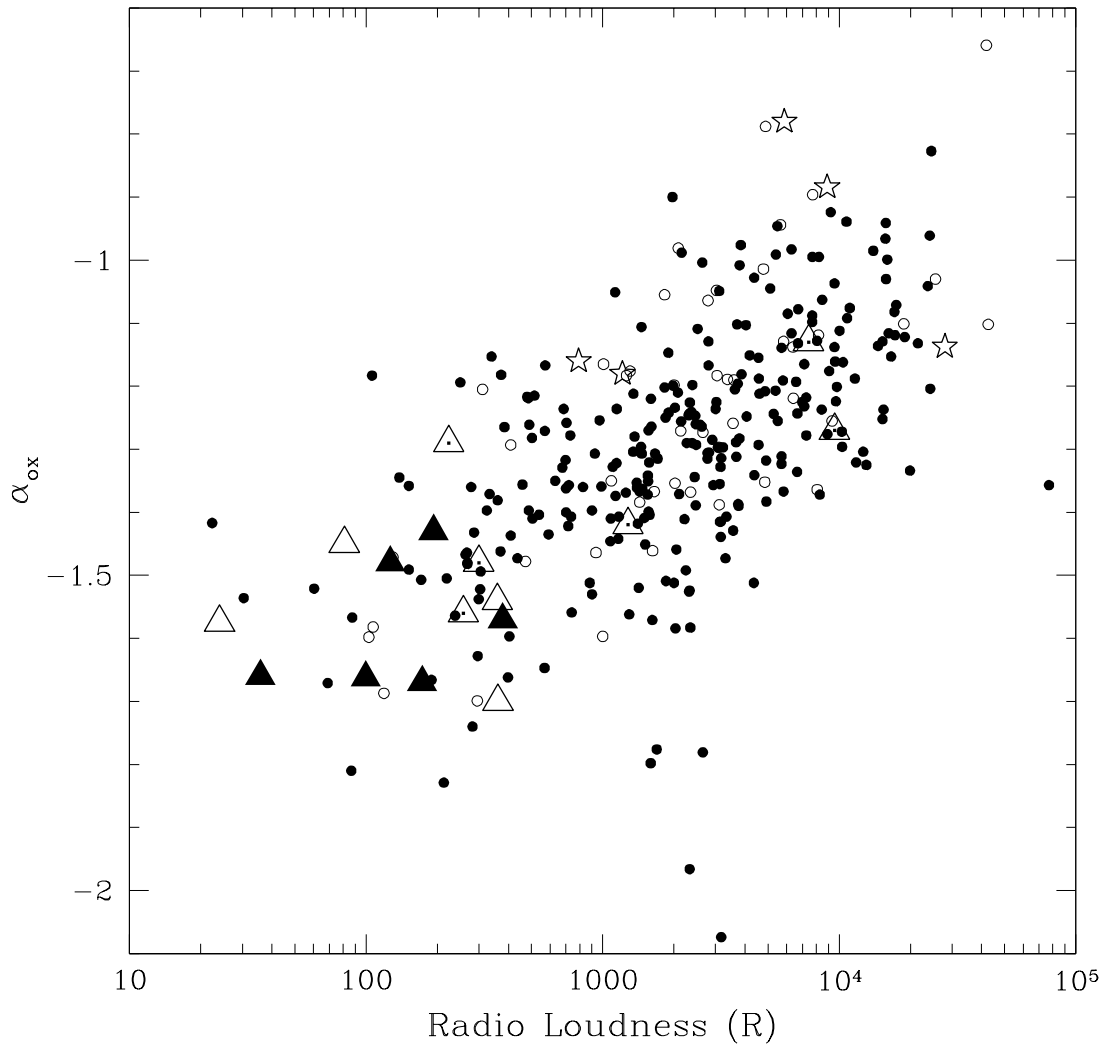}}
\figcaption[f5.eps]{Broad-band spectral index $\alpha_{\rm ox}$ vs. radio loudness ($R$) for samples of flat-spectrum RLQs at different redshifts. The small circles represent the RLQs from the radio-selected sample of Brinkmann et al. (1997) at redshifts of $z$ $\approx$ 0.2--2 ({\it filled circles}) and $z$ $\approx$ 2--4 ({\it open circles}). The large symbols represent moderate-$R$ high-redshift RLQs ({\it triangles}) and blazars ({\it stars}) with X-ray detections. Arrows show upper limits for radio loudness at the $\approx 3\sigma$~level based on the
1.4~GHz radio flux density. In the B04 ({\it filled triangles}) and current ({\it triangles with central dots}) samples, we do not observe any significant X-ray enhancement compared to local RLQs.}
\end{figure*}

\clearpage
\begin{table}
\tabletypesize{\footnotesize}
\tablenum{1}
\begin{center}
\caption{{\it Chandra} Observation Log}
\begin{tabular}{lccccccc}
\tableline
\tableline
\multicolumn{1}{c}{Object} & & Optical & Optical & $\Delta_{\rm Opt-X}$\tablenotemark{a} & X-ray & Exp.~Time$^{\ \rm b}$ \\
\multicolumn{1}{c}{Name} & $z$ & $\alpha_{2000}$ & $\delta_{2000}$ & (arcsec) &
Obs.~Date & (ks) \\
\tableline
PMN~J0214$-$0518 & 3.986 & 02 14 29.30 & $-$05 17 44.6 & 0.16 & 2003 Nov 26 & 4.0 \\
PMN~J0235$-$1805 & 4.314 & 02 34 55.14 & $-$18 06 08.5 & 0.17 & 2004 Feb 07 & 3.9 \\
PMN~J0324$-$2918 & 4.630 & 03 24 44.28 & $-$29 18 21.1 & 0.17 & 2003 Dec 16& 3.8 \\
PMN~J1230$-$1139 & 3.528 & 12 30 55.56 & $-$11 39 09.8 & 0.26 & 2004 Feb 02 & 7.5 \\
PMN~J1451$-$1512 & 4.763 & 14 51 47.05 & $-$15 12 20.0 & 0.21 & 2004 Feb 28 & 4.1 \\
PMN~J2219$-$2719 & 3.634 & 22 19 35.30 & $-$27 19 02.8 & 0.21 & 2003 Nov 19 & 8.1 \\
\tableline
\end{tabular}
\vskip 2pt
\tablecomments{Optical positions of the quasars presented here
can be found in Table 2 of Hook et al. (2002). Optical core positions are used instead of radio core positions because of possible shifts in apparent radio core location from extended radio jets. For PMN J1230$-$1139, we used the precise VLBA radio core position from Beasley et al. (2002) because of possible astrometric errors in the optical position (see $\S$3.2). Units of right ascension are hours, minutes, and seconds, and units of declination are degrees, arcminutes, and arcseconds.}
\tablenotetext{\rm a}{Distance between the optical and X-ray positions.}
\tablenotetext{\rm b}{The {\it Chandra} exposure time has been corrected for detector dead time.}
\end{center}
\end{table}

\begin{table}
\begin{center}
\tablenum{2}
\caption{X-ray Counts, Hardness Ratios, Band Ratios, and Effective Photon Indices}
\tabletypesize{\footnotesize}
\setlength{\tabcolsep}{0.02in}
\begin{tabular}{lccccccc}
\tableline
\tableline
  & \multicolumn{4}{c}{X-ray Counts\tablenotemark{\rm a}} \\
\cline{2-5} \\
Object & [0.3--0.5~keV] & [0.5--2~keV] & [2--8~keV] & [0.5--8~keV] & Hardness Ratio\tablenotemark{\rm b} & Band Ratio\tablenotemark{\rm b} & $\Gamma$\tablenotemark{\rm c} \\
\tableline
PMN J0214$-$0518 & 1.9$^{+2.6}_{-1.2}$ & 32.5$^{+6.7}_{-5.6}$ & 8.8$^{+4.1}_{-2.9}$ & 41.3$^{+7.5}_{-6.4}$ & $-$0.57$^{+0.18}_{-0.14}$ & 0.27$^{+0.16}_{-0.11}$ & 1.8$\pm$0.4 \\
PMN J0235$-$1805 & 2.0$^{+2.6}_{-1.3}$ & 16.7$^{+5.1}_{-4.0}$ & 1.9$^{+2.6}_{-1.2}$ & 20.6$^{+5.5}_{-4.3}$ & $-$0.80$^{+0.22}_{-0.16}$ & 0.11$^{+0.16}_{-0.08}$ & 2.6$^{+1.0}_{-0.8}$ \\
PMN J0324$-$2918 & 2.0$^{+2.6}_{-1.3}$ & 52.0$^{+8.3}_{-7.3}$ & 14.7$^{+4.9}_{-3.8}$ & 66.6$^{+9.3}_{-8.3}$ & $-$0.56$^{+0.18}_{-0.14}$ & 0.28$^{+0.16}_{-0.12}$ & 1.8$\pm$0.4 \\
PMN J1230$-$1139 & 9.0$^{+4.1}_{-2.9}$ & 68.8$^{+9.5}_{-8.4}$ & 31.3$^{+6.6}_{-5.5}$ & 99.9$^{+10.9}_{-9.8}$ & $-$0.37$^{+0.11}_{-0.10}$ & 0.45$^{+0.11}_{-0.10}$ & 1.4$\pm$0.2 \\
PMN J1451$-$1512 & 2.0$^{+2.6}_{-1.3}$ & 70.7$^{+9.6}_{-8.5}$ & 33.5$^{+6.8}_{-5.7}$ & 106.2$^{+11.2}_{-10.2}$ & $-$0.36$^{+0.11}_{-0.10}$ & 0.47$^{+0.12}_{-0.11}$ & 1.4$\pm$0.2 \\
PMN J2219$-$2719 & 5.9$^{+3.6}_{-2.4}$ & 144.1$^{+13.0}_{-11.9}$ & 71.6$^{+9.6}_{-8.5}$ & 215.2$^{+15.7}_{-14.7}$ & $-$0.34$^{+0.12}_{-0.10}$ & 0.50$^{+0.13}_{-0.11}$ & 1.3$\pm$0.2 \\
\tableline
\end{tabular}
\tablenotetext{\rm a}{Errors
on the X-ray counts were computed according to Tables~1 and 2 of
Gehrels (1986) and correspond to the 1$\sigma$~level; these were
calculated using Poisson statistics.}
\tablenotetext{\rm b}{Errors on the hardness ratios [defined as
$(H-S)/(H+S)$, where $S$ is the soft-band counts and
$H$ is the hard-band counts], the band ratios ($H/S$),
and the effective photon indices are at the
$\approx$~$1\sigma$~level and have been computed following the
``numerical method'' described in $\S$~1.7.3 of Lyons (1991). This
avoids the failure of the standard approximate variance formula
when the number of counts is small (see $\S$~2.4.5 of Eadie et al.
1971).}
\tablenotetext{\rm c}{The effective photon indices (and errors) have been
computed from the band ratios and their respective errors, using
the {\sc PIMMS} software.} 
\end{center}
\end{table}
\normalsize

\begin{deluxetable}{lcccccccccccccc}
\pagestyle{empty}
\rotate \tablenum{3} \tablecolumns{15}
\tabletypesize{\tiny}
\setlength{\tabcolsep}{0.02in}
\renewcommand{\arraystretch}{0.6}
\tablewidth{0pt}
\tablecaption{X-ray, Optical, and Radio Properties of Surveyed
Radio-Loud Quasars at High Redshift}
\tablehead{\colhead{Object} & \colhead{$z$} & \colhead{$N_{H}$\tablenotemark{a}~~} & \colhead{$AB_{1450(1+z)}$} &
\colhead{$f_{2500}$\tablenotemark{b}~} & \colhead{$\log (\nu L_\nu
)_{2500}$} & \colhead{$M_B$} &
\colhead{Count~rate\tablenotemark{c}} & \colhead{$f_{\rm
x}$\tablenotemark{d}} & \colhead{$f_{\rm 2\
keV}$\tablenotemark{e}} & \colhead{$\log (\nu L_\nu )_{\rm 2\
keV}$} & \colhead{$\log (L_{\rm 2-10~keV})$\tablenotemark{f}} &
\colhead{$\alpha_{\rm ox}$\tablenotemark{g}} &
\colhead{$\alpha_{\rm r}$\tablenotemark{h}} &
\colhead{$R$\tablenotemark{i}} \\
\colhead{(1)} & \colhead{(2)} & \colhead{(3)} & \colhead{(4)} &
\colhead{(5)} & \colhead{(6)} & \colhead{(7)} & \colhead{(8)} &
\colhead{(9)} & \colhead{(10)} & \colhead{(11)} & \colhead{(12)} &
\colhead{(13)} & \colhead{(14)} & \colhead{(15)} 
} 
\startdata 
PMN~J0214$-$0518 & 3.986 & 2.7 & 18.8 & 14.40 & 46.4 & $-$27.8 & 8.13$^{+1.7}_{-1.4}$ & 3.57$^{+0.75}_{-0.62}$ & 2.01$^{+0.42}_{-0.35}$ &
45.2 & 45.5 & $-$1.48$\pm 0.03$ & $+$0.61 & 299.7 \\
PMN~J0235$-$1805 & 4.314 & 2.6 & 18.9 & 13.13 & 46.7 &
$-$27.8 & 4.3$^{+1.3}_{-1.0}$ & 1.88$^{+0.57}_{-0.44}$ & 1.11$^{+0.33}_{-0.26}$ & 45.3 & 45.6 & $-$1.56$^{+0.04}_{-0.05}$ & $-$0.16 & 257.9 \\
PMN~J0324$-$2918 & 4.630 & 1.2  & 18.6 & 17.31 & 46.9 & $-$28.3 & 13.7$^{+2.2}_{-1.9}$ & 5.80$^{+0.93}_{-0.81}$ & 3.54$^{+0.57}_{-0.49}$ & 45.8 & 46.1 & $-$1.42$^{+0.03}_{-0.02}$ & $+$0.30 & 1281.1 \\
PMN~J1230$-$1139 & 3.528 & 3.5 & 20.1 & 4.35 & 46.1 & $-$26.3 & 9.2$^{+1.3}_{-1.1}$ & 4.12$^{+0.58}_{-0.49}$ & 2.16$^{+0.31}_{-0.26}$ & 45.4 & 45.7 &
$-$1.27$\pm 0.02$ & $-$0.19 & 9550.7 \\
PMN~J1451$-$1512 & 4.763 &  7.8 & 19.0 & 11.98 & 46.5 & $-$28.1 &
17.2$^{+2.3}_{-2.1}$ & 8.59$^{+1.15}_{-1.05}$ &
5.33$^{+0.88}_{-0.65}$ & 46.0 & 46.3 & $-$1.29$^{+0.03}_{-0.02}$ &
$+$0.89 & 223.7 \\
PMN~J2219$-$2719 & 3.634 & 1.4 & 20.3 & 3.62 & 46.1 & $-$26.2 & 17.8$^{+1.6}_{-1.3}$ & 7.57$^{+0.68}_{-0.55}$ & 4.04$^{+0.36}_{-0.30}$ &
45.7 & 46.0 & $-$1.13$^{+0.01}_{-0.02}$ & $-$0.27 & 7426.6 \\
\tableline
\enddata
\tablecomments{~  Luminosities are computed using $H_{0}$=70 km
s$^{-1}$ Mpc$^{-1}$, $\Omega_{\rm M}$=0.3, and
$\Omega_{\Lambda}$=0.7.}
\tablenotetext{a}{From Dickey \& Lockman (1990) in units of
$10^{20}$ cm$^{-2}$ .}
\tablenotetext{b}{Rest-frame 2500~\AA\ flux density in units of
$10^{-28}$ erg~cm$^{-2}$~s$^{-1}$~Hz$^{-1}$.}
\tablenotetext{c}{Observed count rate computed in the 0.5--2~keV band in units of $10^{-3}$ counts
s$^{-1}$.}
\tablenotetext{d}{Galactic absorption-corrected flux in the
observed 0.5--2~keV band in units of $10^{-14}$
erg~cm$^{-2}$~s$^{-1}$. These fluxes and the following X-ray
parameters have been corrected for the ACIS quantum-efficiency
decay at low energy.}
\tablenotetext{e}{Rest-frame 2~keV flux density corrected for
Galactic absorption in units of $10^{-31}$
erg~cm$^{-2}$~s$^{-1}$~Hz$^{-1}$.}
\tablenotetext{f}{Rest-frame 2--10~keV luminosity corrected for
Galactic absorption in units of erg~s$^{-1}$.}
\tablenotetext{g}{Errors have been computed following the
``numerical method'' described in $\S$~1.7.3 of Lyons (1991); both
the statistical uncertainties on the X-ray count rates and the
effects of the observed ranges of the X-ray and optical continuum
shapes have been taken into account (see \S3 for details).}
\tablenotetext{h}{Radio power-law slope in the range 1.4--5~GHz
(observed frame), with $f_{\nu}\propto~\nu^{\alpha}$ from Hook et al. (2002). }
\tablenotetext{i}{Radio-loudness parameter, defined as $R$ =
$f_{\rm 5~GHz}/f_{\rm 4400~\mbox{\scriptsize\AA}}$ (rest frame;
e.g., Kellermann et al. 1989).}
\label{tab3}
\end{deluxetable}

\end{document}